# IOTune: A G-states Driver for Elastic Performance of Block Storage

*Tao Lu[1], Ping Huang[2], Xubin He[2], Matthew Welch[3], Steven Gonzales[3] and Ming Zhang[3]*
[1]New Jersey Institute of Technology   [2]Temple University   [3]Virtustream

## Abstract

Imagining a disk which provides baseline performance at a relatively low price during low-load periods, but when workloads demand more resources, the disk performance is automatically promoted in situ and in real time. In a hardware era, this is hardly achievable. However, this imagined disk is becoming reality due to the technical advances of software-defined storage, which enable volume performance to be adjusted on the fly. We propose *IOTune*, a resource management middleware which employs software-defined storage primitives to implement *G-states* of virtual block devices. G-states enable virtual block devices to serve at multiple performance gears, getting rid of conflicts between immutable resource reservation and dynamic resource demands, and always achieving resource right-provisioning for workloads. Accompanying G-states, we also propose a new block storage pricing policy for cloud providers. Our case study for applying G-states to cloud block storage verifies the effectiveness of the IOTune framework. Trace-replay based evaluations demonstrate that storage volumes with G-states adapt to workload fluctuations. For tenants, G-states enable volumes to provide much better QoS with a same cost of ownership, comparing with static IOPS provisioning and the I/O credit mechanism. G-states also reduce I/O tail latencies by one to two orders of magnitude. From the standpoint of cloud providers, G-states promote storage utilization, creating values and benefiting competitiveness. G-states supported by IOTune provide a new paradigm for storage resource management and pricing in multi-tenant clouds.

## 1 Introduction

Virtualization enables statistical multiplexing of computing, storage and communication devices, which are necessary to achieve illusion of infinite capacity of cloud resources. Persistent states of virtual machines (VMs) are saved in virtual disks, which are image files or logic volumes on physical servers. Virtual disks are commonly hosted by backend storage appliances, which are shared by multiple VMs. Thus, I/O contentions are common occurrences, causing tenants to experience inconsistent storage performance [1, 2, 3, 4, 5].

Hypervisors including QEMU [6], VMWare ESX [7], VirtualBox [8], and Microsoft Hyper-V [9] have implemented functions of limiting disk IOPS or bandwidth, which is also called I/O throttling, to achieve storage performance isolation. The I/O limit defines the resource consumption of a storage volume, thus it is also used to price the volume in public clouds. For example, Amazon EBS *Provisioned IOPS SSD (io1)* volumes charge $0.065 per IOPS-month of provisioned performance with an additional $0.125 per GB-month of provisioned space [10]. Thus, for a 100 GB volume with a provisioned IOPS of 5000, a tenant pays $12.5 for storage space and $325 for storage performance per month. Therefore, performance charge dominates the total cost of using the cloud storage. One common problem of existing cloud block storage is that volume IOPS provisioning is static and immutable after volume creation, which causes two-fold disadvantages. First, the static provisioning cannot adapt to the workload variability and unpredictability. Production workloads are bursty and fluctuant [11, 12, 13, 14, 15, 16]. Peak I/O rates are usually more than one order of magnitude higher than average rates. Thus, static IOPS provisioning places tenants in a dilemma. Under-provisioning fails to support peak loads, resulting in significant I/O tail latencies. Therefore, enterprises are frequently forced to provision workloads with some multiple of their routine load, such as 5-10x the average demand [17]. Over-provisioning may meet peak requirements but wastes a lot of reservation in low-load periods, resulting in extortionate costs. Second, static provisioning loses the resource multiplexing chances, thus, wasting performance capabilities of underlying devices.

To avoid over-provisioning of the computation resource or power consumption under load fluctuations, P-

states and C-states [18] were implemented in processors or co-processors to support multiple performance and energy levels. To mitigate CPU performance interference effects, Q-Clouds dynamically provisions underutilized resources to enable elevated QoS levels, allowing applications to specify multi-level Q-states [19]. Allowing tenants to dynamically update minimum network bandwidth guarantee has also been recognized as useful and critical [20]. However, the feature of multiple levels of QoS has not been achieved at storage device level.

Software-defined storage [21, 3, 4] enables programmable, flexible, and in-situ storage re-provisioning, which is a promising approach to multiple QoS states of storage volumes, achieving resource right-provisioning. IOFlow architecture [21] has demonstrated the feasibility of enforcing end-to-end and in-situ storage resource re-provisioning such as dynamically adjusting the bandwidth limit of a storage share. Motivated by multi-level CPU states, we implement a *G-states* driver called *IOTune* for cloud storage to address the storage resource right-provisioning challenge. IOTune utilizes software-defined storage primitives to support in-situ multi-gear performance scaling up/down for cloud storage volumes.

As a case study, we build SSD storage volumes with elastic performance based on the G-states support of our IOTune framework. Being different from existing *provisioned IOPS* SSD volumes in public clouds [10], which adopt static resource provisioning, volumes with G-states exploit IOPS statistical multiplexing of co-locating volumes and the in-situ IOPS adjustment function supported by software-defined storage primitives to reclaim unused IOPS reservations of underloaded volumes for IOPS promotions of overloaded volumes. The IOPS of a volume can be promoted to serve I/O bursts and demoted thereafter to reduce costs. Our trace-replay based evaluations demonstrate that G-states enable volumes to provide much better QoS with a same cost of ownership, comparing with static IOPS provisioning and the I/O credit mechanism. G-states also reduce I/O tail latencies by one to two orders of magnitude.

In general, G-states supported by IOTune bring three-fold benefits. First, G-states enable storage volume performance to adapt to workload fluctuations. Second, G-states lower price-performance ratio and reduce cost of ownership of cloud storage volumes due to the mitigation of resource over-provisioning. Third, the resource statistical complexing exploited by G-states promotes the utilization of shared storage resources. To summarize, we make the following contributions.

1. We analyze realistic storage traces and recognize the dynamic demand challenge and the statistical complexing opportunity. These insights motivate us to design a resource management framework that enables elastic storage performance.

Table 1: SSD volume IOPS features supported by mainstream IaaS platforms including Google Compute Engine (GCE), Amazon Elastic Block Store (EBS), and Microsoft Azure.

| Platform | SSD volume (128GB) IOPS features | | |
|---|---|---|---|
| | IOPS | Configurable | Change after create |
| GCE [22] | 3840 | No | No |
| EBS io1 [10] | 100-6400 | Yes | No |
| EBS gp2 [10] | 384-3000 | No | Yes |
| Azure [23] | 500 | No | No |

2. We design IOTune, a G-states driver that enables multi-gear elastic performance of block storage. The performance caps enforced by G-states forbid volumes to consume excessive resources, thus ensuring performance isolation among volumes. The resource statistical multiplexing exploited by G-states promotes storage resource utilization.

3. We propose a multi-level pricing model for new storage volumes with G-states. The new price model lowers price-performance ratio of storage volumes without decreasing revenues of providers due to increased resource utilization, thus, creating values for both providers and tenants.

The remainder of this paper is organized as follows. Section 2 presents our problem statement and motivation. Section 3 presents the design of IOTune framework and how to achieve G-states of block storage with IOTune. Section 4 demonstrates how G-states can be applied to lower volume price-performance ratio, meanwhile, to promote storage utilization. Section 5 acknowledges related work. Finally, we conclude the paper.

## 2 Motivation and Problem Statement

We seek for settling the conflicts between immutable resource reservations and dynamic resource demands. Our work was motivated by cloud storage provider Virustream's requirements that storage in multi-tenant environments should have performance isolation, elasticity, and high utilization features. In this section, we first illustrate the resource right-provisioning challenge. Then, we explain the chances that can be taken advantage of to achieve resource right-provisioning. Finally, we explain why G-states is a practical solution, as well as what is required to implement G-states.

### 2.1 A dilemma: fixed reservation vs. dynamic demands

SSD volumes have become an important type of persistent data storage on IaaS platforms [10, 22, 23]. IOPS is a main QoS metric of SSD volumes. Tenants may explicitly [10] or implicitly [22, 23] specify IOPS of SSD volumes. Table 1 lists the SSD volume IOPS features of mainstream IaaS platforms. EC2 EBS *io*1 SSD allows tenants to specify the IOPS of a volume at its creation time. GCE allocates IOPS based on a predefined IOPS to GB ratio, which is currently 30. Azure only provides premium storage disks in three sizes: 128, 512, and 1024GB with predefined IOPS of 500, 2300, and 5000. None of the above enables adjusting the IOPS of a volume after its creation. One exception is EC2 EBS *gp2* SSD, which allows volume IOPS to be promoted up to 3000 during bursts using the I/O credit mechanism [10].

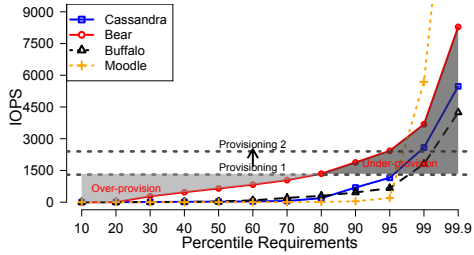

Figure 1: IOPS requirements of real workloads. Peak loads are much higher than the average.

Table 2: IOPS distributions of six storage volumes. Each volume backs a one-hour episode of the Bear trace.

| Volume | IOPS | | | | |
|---|---|---|---|---|---|
| | Average | 90% | 95% | 99% | 99.9% |
| 1 | 906 | 1877 | 3255 | 4026 | 5592 |
| 2 | 632 | 1626 | 2289 | 4433 | 6976 |
| 3 | 338 | 1084 | 1412 | 2050 | 3271 |
| 4 | 362 | 1077 | 1439 | 2192 | 2739 |
| 5 | 396 | 1257 | 1570 | 3262 | 6940 |
| 6 | 347 | 1121 | 1390 | 2024 | 5133 |
| Sum | 2981 | 8042 | 11355 | 17987 | 30651 |
| Multiplex | 2981 | 6793 | 7966 | 10387 | 13469 |

Handling workloads bursts is particularly challenging in IOPS static reservation of storage volumes. To reveal the demand dynamics, we analyze the IOPS requirements of real workloads using four traces: Bear, Buffalo, Moodle [24], and Cassandra. Cassandra traces contain I/O statistics of three production Cassandra VMs on Virtustream's public cloud platforms. The percentile IOPS requirements are shown in Figure 1. One common characteristic of these workloads is that the volumes have low or moderate IOPS requirements in more than 70% of the time. However, the tail IOPS requirements exponentially hike. Our analysis on Bear trace shows that top 30% peak periods contribute about 70% of the total I/O requests.

Assuming we need to create a *provisioned IOPS* SSD volume for Bear workloads and we expect that 80% of the time the IOPS requirement can be satisfied. We have to provision the volume IOPS as *provisioning 1*, which is 1300, in the figure. This is a moderate provisioning. However, in the other 20% of the time the IOPS is under-provisioned and workloads will notice long response time. As an alternative, the IOPS can be set as *provisioning 2*, which is 2400, satisfying 95 percentile requirements. However, the resource over-provisioning will cause serious waste during the low-load periods. In general, IOPS static reservation is not able to cope with workload fluctuations and rarely achieves resource right-provisioning. Although I/O bursting mechanism [10] can alleviate this conflict in certain circumstances, however, during continuous I/O bursts, volumes can hardly accumulate enough credit balance, thus, the bursting mechanism regresses to static reservation.

## 2.2 Chances: IOPS statistical multiplexing and software-defined storage

SRCMap [13] recognized significant variability in I/O intensity on storage volumes. Everest [12] validated the chance of statistical multiplexing of storage resources in production environments. We briefly demonstrate the chance of IOPS statistical multiplexing of co-locating volumes. We adopt the approach of concurrently replaying six one-hour trace episodes[1] on six different SSD volumes. Table 2 lists the percentile IOPS of the volumes and their statistical aggregates. The aggregate peak I/O rate is obviously lower than the sum of each individual peak I/O rate. For example, the sum of the 95th% IOPS of all episodes is 11355 while the 95th% aggregate IOPS in multiplexing is only 7966, which is 30% less than the sum due to stagger I/O peaks of volumes. Assuming the six volumes are all provisioned with IOPS of their 90th% arrival rates, the total IOPS reservation will be 8042, satisfying the 95th% aggregate IOPS requirement if the IOPS reservations of all volumes are multiplexed.

## 2.3 G-states: A promising solution

I/O credit mechanism is a state-of-the-art solution for satisfying I/O bursts [10]. The leaky bucket algorithm [25], which was originally designed as a flow control mechanism for ATM network, is the core of I/O credit mechanism. The basic idea of the algorithm is when I/O demand drops below the baseline level, unused credits are added to the I/O credit balance, which will be consumed in the future to sustain burst I/O performance. I/O credit mechanism has two main limitations. First, I/O credit does not take the underlying storage device utilization into consideration, thus, may result in improper resource allocation decisions. With I/O credit mechanism, if the credit balance of a volume runs out and the volume has I/O bursts, no promoted performance will be offered to the volume, even if the underlying storage system ac-tually has spare resources. This will cause suboptimal device utilization, since the spare capability can be allocated to the volume without impacting other volumes. Second, the credit accumulation may take a long time. For example, a volume with a baseline IOPS of 300 takes at least ten seconds to accumulate credit balance of 3000, which can only serve a burst period as long as one second. Therefore, if the I/O bursts are relatively intensive, I/O credit mechanism does not work well. The performance of the leaky bucket algorithm is arrival pattern dependent, which has been reported in [26].

---
[1]We adopt this method because there lacks publicly available concurrent block I/O traces of co-locating volumes. The exact subtrace we use is from http://visa.lab.asu.edu/traces/bear/blkios-2012323010000.gz

For computation performance, processor P-states, C-states, and application Q-states [19] have demonstrated the feasibility and effectiveness of multi-level QoS-aware resource re-provisioning. Q-Clouds [19] tunes resource allocations to mitigate performance interference effects, dynamically provisioning underutilized resources to enable elevated QoS levels, which is called application Q-states, thereby improving system efficiency. Storage resource faces a right-provisioning challenge. Since current cloud storage already provides a baseline performance for a volume. Enabling multi-level performance will be a natural extension of the existing performance and pricing model of cloud resource. This motivates us to design and implement a multi-level resource re-provisioning control framework for cloud storage.

We propose IOTune, a G-states driver for block storage. We use the term ***G-states to define capability states of virtual block devices***. While a virtual block device is in $G0$ performance state, it uses its baseline performance capability, which is designed to be specified by the tenant at volume creation, requiring minimum resource reservation. When workloads demand more capability than the current level, the G-state of the block device is promoted by one level, its performance capability is doubled, requiring doubled resource reservation. $Gn(n \geq 0)$ performance state bears $2^n$ times performance capability of $G0$. In public clouds, we expect G0 to be a provider guaranteed QoS level, while other performance gears are best-effort, depending on available resources. G-states cap the resource consumption of a volume, enable cloud storage to have multi-level performance elasticities, and promote storage resource utilization.

The G-states supported by IOTune framework shares the same philosophy with previous multi-states mechanisms. However, the resource allocation of storage system has very different design concerns. We present the IOTune design and implementation in details, demonstrating the challenges we have overcome.

## 3 IOTune

In this section, we introduce the basic elements, design, and architecture of IOTune. We discuss the position of IOTune in the virtualization system, the working process of IOTune, the run-time system information required for IOTune to make resource re-provisioning decision, and the interaction of IOTune with other system components. We focus on discussing how G-states of storage volumes are supported with IOTune.

Table 3: IOTune Building Blocks: Libvirt Virtualization Primitives for Block Device Management.

| Primitive | Function |
|---|---|
| 1. blkdeviotune | Tune device total storage bandwidth  <br>Tune device read bandwidth  <br>Tune device write bandwidth  <br>Tune device total storage IOPS  <br>Tune device read IOPS  <br>Tune device write IOPS |
| 2. blkiotune | Tune storage shares of VMs  <br>Tune storage shares of devices within a VM |
| 3. domblkstat | Obtain real time block device I/O statistics |

### 3.1 Building Blocks

IOTune is implemented in user space. IOTune interacts with hypervisors, utilizing simple software-defined storage primitives provided by libvirt virtualization API [27] to achieve elastic storage resource re-provisioning. Virtualization primitives for block device management are the building blocks for our IOTune framework. Table 3 summarizes current virtualization primitives for performance tuning of storage volumes, tuning can be committed at block device level or VM level. Hypervisors also provide I/O statistics that are critical for tuning. For example, *blkdeviotune* with an IOPS parameter can adjust the IOPS performance of a target storage volume. And this adjustment is in situ and in real time, enabling IOTune to implement G-states of block storage volumes.

### 3.2 Design Overview

IOTune is designed as a middleware for Infrastructure as a Service (IaaS) platforms. It supports system-level configurations such as baseline IOPS per GB, aggregate performance limits, utilization threshold of physical device for I/O tuning, unit IOPS per GB price, and so on. Tuning decisions are also made based on these parameters. The IOTune execution procedure includes two stages.

**Stage 1: Volume Instantiation.** A volume is a block device management unit, which together with capacity and IOPS bills forms a billing entity. Therefore, a volume is a natural management unit in the IOTune framework. When creating a storage volume, the tenant specifies a requested size of the volume. Once the creation completes, IOTune instantiates the volume, pulls volume information including storage path, size, creation time, calculates the multi-level IOPS metrics, and initializes the metering data.

**Stage 2: Continuous I/O Tuning.** Upon the instanti-

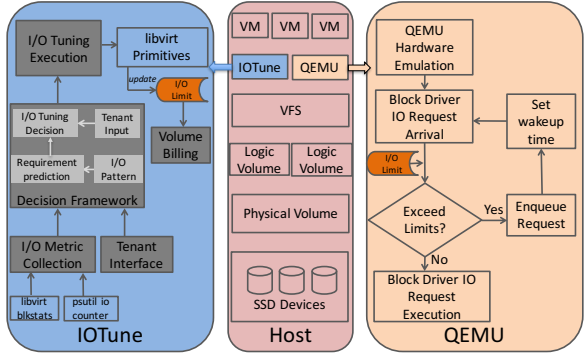

Figure 2: A systematic overview of IOTune in the QEMU/KVM hypervisor based virtualization system. IOTune is a userspace program utilizing software-defined storage primitives of QEMU to implement G-states of block storage volumes.

ation of a volume, IOTune conducts initial tuning, which sets a baseline IOPS limit on the volume. Then, IOTune periodically makes tuning decisions. By default, we set the tuning interval as one second, existing work [15] as well as our trace analysis show that a lot of I/O bursts last only for a few seconds. We believe a fine granularity tuning is necessary to timely satisfy the resource demands of these short I/O bursts. For IOPS resource management, IOTune checks the real-time IOPS monitoring data to judge the promotion or demotion of current IOPS level. Once tuning decisions are made, IOTune calls virtualization primitives listed in Table 3 to make real-time in-situ performance resource re-provisioning to implement G-states, achieving elasticities of storage volumes.

### 3.3 Architecture

Figure 2 presents a systematic overview of IOTune in the QEMU/KVM hypervisor based virtualization system, as well as its interactions with other system components. The key objective of IOTune is in-situ adjusting volume resource allocations based on real-time demands of workloads. To achieve this, IOTune deploys a module for collecting I/O statistics, a module for making tuning decisions, and a module for enforcing tuning. The decision making depends on requirement prediction, which can be based on historical statistics or real-time demands. In current implementation, we compare the real-time IOPS with the reserved IOPS of a volume to judge whether it is overloaded. The performance gear of a volume can be promoted only if underlying devices have spare capability. Device level I/O metrics are required to calculate the storage utilization. Once a tuning decision is made, the resource re-provisioning command is committed to the volume. Metering and pricing policies are also required to calculate the bill of a volume in a billing period. In general, the volume

---

**Algorithm 1:** `IOTune`: Adaptive I/O Tuning

**Input:** $V_i$ (i = 0, 1, ..., M): Logic volumes;
T: I/O type;

```
// ① Get initial multi-level IOPS
   settings of volumes
```
**for** i = 0, 1, ..., M **do**
  $Gears_i$ ← getiopsgears($V_i$, T)
**end**
```
// ② Tune volumes continuously.
   Tuning period is one second
```
**for** *Every tuning period* **do**
  **for** *Each $V_i$, i = 0, 1, ..., M* **do**
    ```
    // ③ Make tuning decision
    ```
    $Tune_i(t)$ ← TuneJudge($V_i$, T, $Gears_i$, $Threshold_i$)
    **if** $Tune_i(t)$ *is "promote"* **then**
      ```
      // ④ Commit promotion to
         virtual volumes, promote
         IOPS gear by one level
      ```
      TuneExecute($V_i$, $Tune_i(t)$, T)
      PromoteIOPS($V_i$, T)
    **end**
    **if** $Tune_i(t)$ *is "demote"* **then**
      ```
      // ⑤ Commit demotion to
         virtual volumes, demote
         IOPS gear by one level
      ```
      TuneExecute($V_i$, $Tune_i(t)$, T)
      DemoteIOPS($V_i$, T)
    **end**
  **end**
**end**

---

management function of IOTune framework is supported by I/O monitoring, tuning decision, tuning execution, and metering modules. The detailed working procedures of IOTune are demonstrated in Algorithm 1.

**I/O Monitoring.** IOTune collects two categories of I/O metrics. The first is the I/O metrics of virtual disks. On our platforms, QEMU block driver layer provides interfaces to obtain the I/O statistics of each virtual disk. Considering QEMU block driver is also the location where resource provisioning is enforced, IOTune collects virtual disk I/O statistics and conducts resource provisioning at QEMU block driver layer. IOTune employs *libvirt* API which calls *DomainBlockStats* interface of QEMU to obtain the IOPS and bandwidth data of target virtual disks. The second category of I/O metrics relating to physical devices. The decision making framework of IOTune takes the physical device utilization into consideration. I/O monitoring component of IOTune reads block device I/O counters via

**Algorithm 2:** `StorageUtil`: Get the performance utilization of a physical volume

**Input:** dev: Physical volume;
  MaxRIOPS: Physical volume read IOPS limit;
  MaxWIOPS: Physical volume write IOPS limit;
  MaxRBW: Physical volume read bandwidth limit;
  MaxWBW: Physical volume write bandwidth limit;
**Output:** Real-time utilization of a physical volume;
```
// ① Collect I/O metrics via
    psutil API
```
  riops, wiops, rbw, wbw ← metricvalues(dev)
```
// ② Calculate IOPS utilization
```
  iopsutil ← riops / MaxRIOPS + wiops / MaxWIOPS
```
// ③ Calculate bandwidth
    utilization
```
  bwutil ← rbw / MaxRBW + wbw / MaxWBW
```
// ④ Return physical volume
    utilization
```
  return max(iopsutil, bwutil)

---

*psutil* API to calculate the IOPS and bandwidth usage of underlying physical volumes. These metrics are further used to calculate storage performance utilization. It is challenging to accurately estimate the storage capability utilization [28], especially for devices serving requests in parallel such as RAID arrays comprising multiple devices and modern SSDs containing multiple I/O channels [29]. Our underlying storage employs RAID5 SSD arrays, therefore, traditional storage utilization monitoring utilities such as the widely used *iostat* do not work well. We calculate the storage device utilization based on offline evaluations. We beforehand measure the maximum read/write IOPS and bandwidth of the RAID array under various thread numbers. To calculate the real-time physical device utilization, we collect the real-time IOPS and bandwidth statistics, differentiating read and write to separately calculate the utilization in IOPS and bandwidth dimensions. The higher utilization of IOPS and bandwidth represents the underlying device utilization. The calculation of storage utilization is presented in Algorithm 2.

**Decision Making.** IOTune decision module decides *when* to adjust the IOPS of *which* volume by *how much*. Temporal I/O patterns can be used to predict the volume IOPS requirements so as to promote the IOPS level of a volume before bursts arrive. For example, diurnal variations have been recognized in web server [11], Hotmail and Messenger [14] disk I/O loads, so even the wall-clock time can be a hint for predicting volume resource requirements. These statistical patterns can be useful for coarse-grained tuning. However, G-states of storage volumes require real-time and accurate tuning, which can hardly be achieved by prediction. I/O Monitoring module of IOTune collects real-time volume performance metrics including IOPS and bandwidth, which reflect real-time resource demands of a volume.

Most of the time I/O requirements of all volumes can be satisfied because I/O peaks are stagger. But in extreme cases, peaks of volumes may overlap. In such promotion contention scenarios, fairness and efficiency are two considerations for making tuning decisions. For fairness, the promotion of volume which has the lowest current IOPS level should be prioritized. For efficiency, the promotions which will maximize storage utilization should be applied. Since IOTune is designed to run at cloud provider side, we believe it is more reasonable to adopt the efficiency first policy so that providers can get more revenues. In promotion contention scenario, IOTune will promote the performance level of a volume so that the storage utilization can be maximum.

Considering the arriving and fading of I/O peaks are both immediate, IOTune employs multiplicative increase and decrease to adjust volume resource reservations. Specifically, volume IOPS is doubled for promotion, and is halved for demotion. This policy also simplifies the reservation metering and the pricing policy.

For workloads that don't have a high requirement for quick response, IOPS promotion may not be attractive. Batch processing is an example. At volume creation, service providers can offer tenants an option to disable the automatic resource adjustment of volumes to avoid bills that are not for business critical workloads.

Algorithm 3 demonstrates IOPS resource management procedures. At volume creation, the volume is initiated at its baseline performance gear. Every second, IOTune checks the real-time IOPS value and level of the volume, as well as calculates the storage utilization. The tuning decision is made as follows. If the IOPS value reaches the cap of current gear, the current gear is not the top gear, and the storage utilization does not reach the threshold, the IOPS gear of the volume is promoted by one level and the volume performance is doubled. Similarly, if the IOPS consumption of a volume is less than the lower gear limit, its IOPS gear is demoted by one level. Otherwise, the performance gear of the volume does not change.

**Tuning Execution.** Tuning execution uses *libvirt* virtualization primitives listed in Table 3, which call QEMU block device tuning interfaces. VM I/O requests traverse virtualization storage stacks and become host-side asynchronous I/O requests executed by QEMU. For storage performance management, QEMU provides software interfaces for cloud operators to specify the IOPS and bandwidth limits of volumes [6]. As it is demonstrated in Figure 2, upon the arrival of a block driver *aio* request, the QEMU *system emulator block driver* intercepts it and checks if an I/O wait is needed for I/O rate throttling purpose. If an I/O wait is needed, the request will be sent into a QEMU *coroutine queue*, waiting for a specific time determined by a *throttle schedule timer* and then be executed. The main I/O

**Algorithm 3:** `TuneJudge`: Make Tuning Decision
**Input:** $V_i$: Logic volumes;
$Gears_i$: Initial multi-level IOPS settings;
T: I/O type;
$Threshold_i$: Utilization threshold of physical device;
**Output:** Tuning decision;
// ① Get current IOPS level
$Level_i \leftarrow$ getiopslevel($V_i$, T)
// ② Get the real-time IOPS value
$IOPS_i(t) \leftarrow$ getiops($V_i$.lvpath, T)
// ③ Promote: Volume IOPS reaches its limit and current level is not the top level
**if** $IOPS_i(t) > Gears_i[Level_i] * 0.95$ **AND** $Level_i <$ len($Gears_i$) - 1 **then**
  // ④ Get current physical device utilization
  $Util_i(t) \leftarrow$ StorageUtil($V_i$.getpv())
  **if** $Util_i(t) < Threshold_i$ **then**
    $Util_i(t) \leftarrow$ StorageUtil($V_i$.getpv())
    return "promote"
  **end**
**end**
// ⑤ Demote: IOPS demand is less than its lower level limit
**if** $Level_i > 0$ **AND** $IOPS_i(t) < Gears_i[Level_i - 1]$ **then**
  return "demote"
**end**
return None

---

**Algorithm 4:** `TuneExecute`: Commit updated performance parameters to storage volumes
**Input:** $V_i$: Logic volumes;
$TuneType$: promote or demote;
T: I/O type;
// ① Identify target instance and device
instance, blkdev $\leftarrow$ GetTargetDev($V_i$.lvpath)
// ② Commit new performance value to storage volume
libvirt_blkdeviotune(instance, blkdev, T, $V_i$.curiopsvalue, $TuneType$)

---

management feature provided by IOTune is dynamically and in-situ commit resource adjustments of volumes in real time according to the time-variant I/O requirements of the volume and the tuning decisions. The tuning execution procedures are simple. As Algorithm 4 shows, first the target instance and device are identified, then the *libvirt blkdeviotune* primitive is executed with parameters deduced from inputs.

**Volume Pricing Policy.** policy are essential for public clouds, in which storage resources are charged. For an SSD volume, it's capacity and IOPS reservation are charged. This is the practice of cloud providers like Google and Amazon. Currently, Google charges SSD only based on its capacity, while Amazon charges the SSD capacity and IOPS separately [10]. IOTune takes the pricing policy into consideration for designing the elastic storage volumes. IOTune targets the platforms where IOPS capability is charged separately. Since the virtual disk managed by IOTune has multiple performance gears, and the IOPS reservations of a volume is dynamic during its lifetime due to the QoS-aware allocation policy, the pricing policy of IOTune is more complicated. IOTune meters the duration a volume served at each QoS level. Charges at all levels sum up to the total bill. The total bill is calculated as follows:

$$TotalBill = CapacityBill + QoSBill \qquad (1)$$

The total volume bill consists of capacity bill and QoS bill. The capacity bill charges for storage space. The QoS bill charges for performance of the storage volume.

$$CapacityBill = PerGBRate * VolSize * BillPeriod \qquad (2)$$

Capacity bill of a volume is the product of its per GB price, volume size and the billing period time.

$$QoSBill = \sum_{i=0}^{N} QoSBill_i \qquad (3)$$

QoS bill consists of bills a volume serving at all performance gears.

$$QoSBill_i = RateGi * DurationGi \qquad (4)$$

The $Gi$ QoS bill is the product of its $Gi$ price and the duration of the volume served at $Gi$. The $Gi$ prices of various volumes are different, which are related to baseline IOPS or bandwidth specified by tenants. $DurationGi$ is the active time a volume served at $Gi$.

**Storage-specific Issues.** IOTune handles storage-specific issues including the I/O request type and size. Storage systems usually show different read and write performance. This can be caused by device performance features and data layouts. For example, SSDs usually show higher read IOPS than write. Different data layouts on multiple devices for data mirroring or parity may also yield various aggregate read and write performance. IOTune supports separate IOPS tuning for read and write. Larger I/O sizes consume more storage bandwidth. For a volume issuing mostly large requests, instead of IOPS, storage bandwidth will be the dominant bottleneck. Either the provisioned IOPS metric may not be achieved,

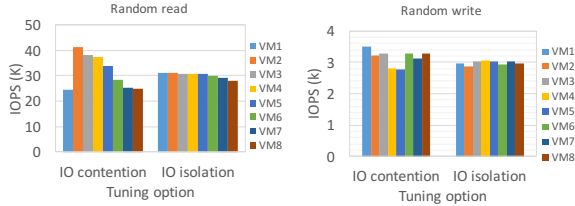

(a) Random read: I/O contention vs. isolation.  (b) Random write: I/O contention vs. isolation.

Figure 3: Performance isolation achieved with virtualization I/O tuning primitives. I/O contention and I/O isolation indicate without and with performance cap applied to virtual disks, respectively.

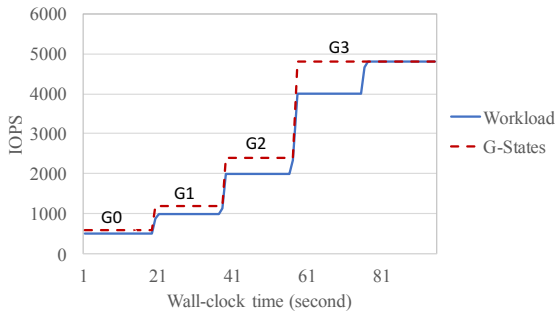

Figure 4: Demonstration of how G-states work.

or the single volume will consume too much bandwidth resource. Thus I/O requests with various sizes should be treated differently. Although we focus on discussing IOPS, in practice IOTune takes bandwidth into consideration even for making IOPS tuning decisions.

## 4 Implementation and Evaluation

We implement IOTune on a QEMU/KVM virtualization storage stack, with it running in the user space of the host machine. IOTune is a cloud provider oriented storage management framework. Cloud providers can make use of IOTune without modifying the host kernel, guest operating systems, or applications. Our current implementation is based on LVM block storage. VM disks are logic volumes on the host machine.

In this section, we present results from a detailed evaluation of IOTune. Our physical machine consists of two Intel(R) Xeon(R) CPU E5-2670 v3 @ 2.30GHz; 16 x 16GB DDR4 DRAM, totaling 256GB; the physical storage devices are 6 x 2TB SATA SSD in RAID5 with a total of 10TB available space. The host OS is a 64-bit Ubuntu 14.04 with Linux kernel version 3.13.0-85-generic and the KVM module. QEMU emulator version is 2.4.0. Ubuntu 14.04 64-bit image is run on the VM as the guest operating system with 4 VCPUs, 16GB memory and ten 100GB SSD virtual disks.

Table 4: Workload resource reservation configurations with IOTune, LeakyBucket, and Static provisioning.

| Workload | IOPS Value | | | | | |
|---|---|---|---|---|---|---|
| | Static | Leaky Bucket | IOTune | | | |
| | | | G0 | G1 | G2 | G3 |
| A | 1100 | 1100 | 600 | 1200 | 2400 | 4800 |
| B | 3000 | 3000 | 1300 | 2600 | 5200 | 10400 |

### 4.1 Evaluation of Virtualization Primitives

Software-defined storage primitives specifying the IOPS and bandwidth of storage volumes are the backbone of IOTune framework. Accuracies of the primitives are crucial to the usefulness of IOTune. To evaluate the primitive accuracy, we execute *blkdeviotune* with –*total_iops_sec* and –*total_bytes_sec* parameters, respectively, on the hypervisor to enforce the IOPS or the bandwidth limit of a storage volume. We run *fio* benchmark in *DIRECTIO* mode on the VM where the volume is mounted. We use single I/O thread with queue depth as one. For IOPS and bandwidth evaluation, we use 4KB and 128KB I/O requests, respectively. We compare the performance metrics reported by *fio*, which is also the tenant-observed performance. For evaluation, we range the IOPS from 100 to 16000 and the bandwidth from 1 to 128 MBps. Our evaluation demonstrates that the virtualization primitive for IOPS enforcement has a deviation of less than 0.3%; the primitive for bandwidth enforcement has a deviation of less than 0.1%, both of which are very accurate.

We also evaluate the effectiveness of software-defined storage primitive for performance isolation in the shared storage scenario. We run eight VMs to share the physical storage. As Figure 3 shows, when the virtual disks compete resources freely without any limitation, their performance variance can be up to 42%. When all virtual disks are set with a same performance cap, their performance variance is under 8% for both read and write. The read performance is much better than write, because SSD devices usually have better read performance than write. Also, RAID5 is excellent in random read but only fair in random write due to parity overhead. The effectiveness of existing software-defined storage primitives lays a good foundation for the usefulness of IOTune.

### 4.2 How G-states Work

We run a simple synthetic *fio* workload on a storage volume to demonstrate how G-states work. The workload consists of five twenty-second phases from phase0 to phase4 with IOPS of 500, 1000, 2000, 4000, and 6000, respectively. G-states of the volume is configured with four gears from G0 to G3 with IOPS of 600, 1200, 2400, and 4800, respectively. The workload run-time throughput is presented in Figure 4.

Workload demands trigger G-states transitions. The volume is initiated with G0. Since G0 bears an IOPS of 600 which is higher than the phase0 demand, the workload in phase0 can achieve its performance. Entering phase1, the workload has a higher IOPS demand of 1000, but the volume IOPS capability is 600, IOTune will notice in real time that the volume performance reaches its current cap. Therefore, the volume performance gear will be promoted to 1200, which is higher than 1000, thus the phase1 demand will be satisfied again. Phase2 and phase3 experience similarly as phase1.

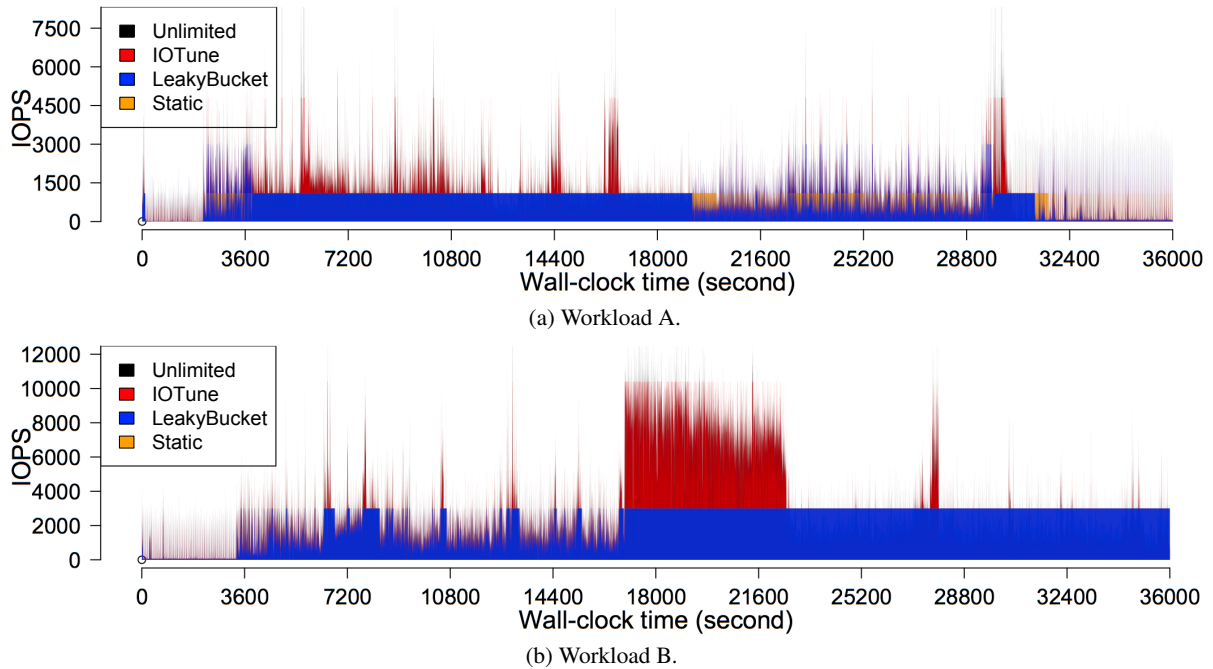

Figure 5: Workload run-time I/O throughput with IOTune, LeakyBucket, and Static resource provisioning, respectively. In the first 4.5 hours, the I/O bursting supported by LeakyBucket promotes I/O throughput. But once the initial credit balance runs out, LeakyBucket works like the Static. The G-states supported by IOTune enable volume resource provisioning always adapts to dynamic workload demands.

G-states throttle workload resource consumption. Entering phase4, the workload demands an IOPS of 6000, but at that time G-states of the volume reach the highest gear, no further gear promotion can be made. Therefore, the workload will be throttled at the IOPS throughput of 4800, which equals the G3 IOPS capability.

## 4.3 The Effectiveness of G-states

In this section we evaluate the effectiveness of G-states for production workloads. IOPS, tail latency, and cost of ownership are key metrics for data center disks [30]. We examine the following key questions: (1) To achieve a same IOPS level how much resource reservation can G-states reduce? (2) For a same volume admission control policy, how much I/O tail latency and storage utilization can G-states improve?

### 4.3.1 Meeting QoS requirement with reduced resource reservation

Our evaluation demonstrates that G-states of IOTune enable storage volumes to meet IOPS requirements with reduced resource reservations. We compare the volume IOPS under various resource provisioning policies via replaying the two Bear subtraces[2] [24]: a 22-hour Workload A, and a 17-hour Workload B. The I/O rate of Workload A is moderate. The I/O rate of Workload B is high. So they represent diverse application features. For legibility purpose, Figure 5 shows the run-time volume IOPS of the first ten hours.

*Unlimited* indicates there is no IOPS limit imposed to the volume, thus representing the natural I/O arrival rates. *Static* is set as the 85 percentile IOPS requirement of workloads, which is 1100 and 3000 for workload A and B respectively. For a fair comparison, we adopt offline calculation to decide IOTune parameters so that considering volume bills IOTune costs as much as Static and LeakyBucket. Workload resource reservation configurations are presented in Table 4. Figure 5 shows the live volume throughputs with IOTune, LeakyBucket, and Static resource allocation policies. For Static, whenever there are I/O bursts that require more resources than the reservation, the excessive requests are seriously delayed. For LeakyBucket policy, two key parameters are *maximum credit balance* and *burst IOPS*. Currently, EBS General Purpose SSD (gp2) volumes are allowed to have a maximum credit balance of 5.4 million and burst IOPS of 3000. The I/O credit accumulation rate is 3 IOPS/GB per second. We use these parameters for our tests of LeakyBucket policy. When there is I/O credit balance, the volume throughput can be temporarily promoted up to 3000 to satisfy the high resource demands. However, once the credit balance runs out, LeakyBucket regresses to the Static and the excessive requests are also seriously delayed. For workload A, in the first hour the I/O intensity is low, so the LeakyBucket can accumulate considerable credit balance to satisfy the occasional I/O bursts. But during the second to the fifth hour when the I/O intensity is moderate, LeakyBucket fails to accumulate enough credit balance to satisfy I/O bursts on time and a lot of requests are delayed. In contrast, although the baseline IOPS reservation of IOTune is set at mere 600 in this case, the 4-gear G-states of storage volumes supported by IOTune enable dynamic IOPS promotion so that the volume can achieve a peak I/O processing rate of 4800 during bursts and always satisfy the occasional high resource demands. For Workload B, since its baseline IOPS is 3000 which is the same as the burst IOPS under LeakyBucket policy, therefore, the I/O bursting mechanism does not help Workload B at all. So for Workload B LeakyBucket works exactly the same as the Static. But IOTune can still utilize its multi-gear mechanism to ensure resource allocation always adapt to the workload fluctuation.

---
[2]Trace file of Workload A: http://visa.lab.asu.edu/traces/bear/blkios-2012323010000.gz; Workload B: http://visa.lab.asu.edu/traces/bear/blkios-2012324010000.gz

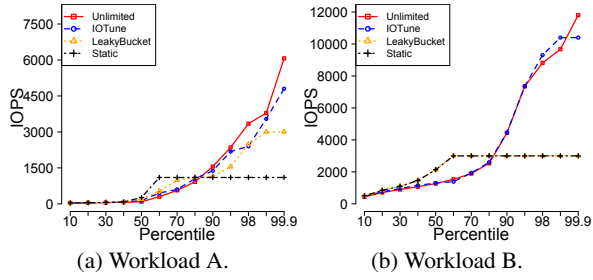

(a) Workload A. (b) Workload B.

Figure 6: Workload IOPS distributions under various provisioning policies.

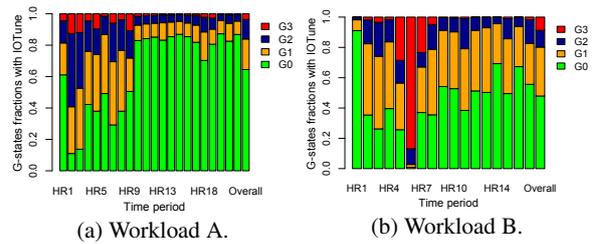

(a) Workload A. (b) Workload B.

Figure 7: The duration of workload served at each G-states level with IOTune.

Figure 6 demonstrates quantitative comparisons of the IOPS distributions under all policies. IOTune enables volumes to achieve near optimal performance, as if it has been provisioned with unlimited performance, in more than 95% of the time. In the other 5% of the time, IOTune also enables workload to achieves at least 80% throughput of the Unlimited case. For Workload A, Static provisioning only satisfies the 85 percentile I/O request rate, and about one half of the I/O requests are seriously delayed. LeakyBucket also only satisfies the 85 percentile I/O request rate, but compared with Static the I/O bursting of LeakyBucket can reduce the queuing latencies of delayed requests. For Workload B, both Static and LeakyBucket provisioning only satisfy the 85 percentile I/O request rate. IOTune satisfies the 99 percentile I/O request rate. Even for 99.9 percentile I/O request rate, IOTune achieves more than 80% of the unlimited case.

The Static provisioning policy has constant resource reservation. IOTune adjusts resource reservations on the fly in four gears: G0, G1, G2 and G3. IOTune writes IOPS logs so we can calculate the durations the volume is served at each IOPS level. As it is demonstrated in Figure 7, in more than 80% of the time volumes serve at low or moderate resource reservation level, G0 and G1. The reservation is promoted only when workloads demand more performance quota. For Workload A, only during the second and the third hour G2 and G3 account for more than 40% of the time. For Workload B, only during the fifth and the sixth hour G2 and G3 account for more than 40% of the time. We use the pricing policy of EBS Provisioned IOPS SSD (io1) volume, $0.065 per provisioned IOPS-month, to compare the IOPS bills of IOTune with the Static and LeakyBucket policies. From Figure 8 we can see that in 15 out of 22 hours of Workload A, and 13 out of 17 hours of Workload B, IOTune costs less than the Static or LeakyBucket policies. And the total IOPS bills of IOTune is $2.20 for Workload A, and $4.77 for Workload B; the total bill for Static is $2.18 for Workload A, and $4.60 for Workload B. LeakyBucket costs the same as Static. Although costing about the same, IOTune performance is much better than Static or LeakyBucket in general, particularly in I/O intensive periods. The Static policy has to double its reservation to achieve a comparable performance as of IOTune.

### 4.3.2 Improving end-to-end I/O latency and storage utilization

Our evaluation also demonstrates that IOTune reduces end-to-end I/O latencies and improve storage utilization. Short end-to-end I/O latency is attractive to tenants. It is also a metric to measure the QoS of storage. Storage uti-lization metric summarizes how well a provider manages its storage assets across the entire business. Achieving high storage utilization is important for cloud providers to save costs and to promote competitiveness. We replay the six traces of Table 2 each on a 100GB SSD virtual disk. We set the static IOPS limit of each volume at the 90th% requirement of its workload, so the total IOPS reservation for the six volumes is 8047. For IOTune, we ensure the same total IOPS reservation. The baseline G0 IOPS of each volume is the same as the Static reservation case. For a fair comparison, if the IOPS of a volume needs to be promoted in IOTune case, the promotion can be executed only if the unused total reservation is more than the promotion requirement.

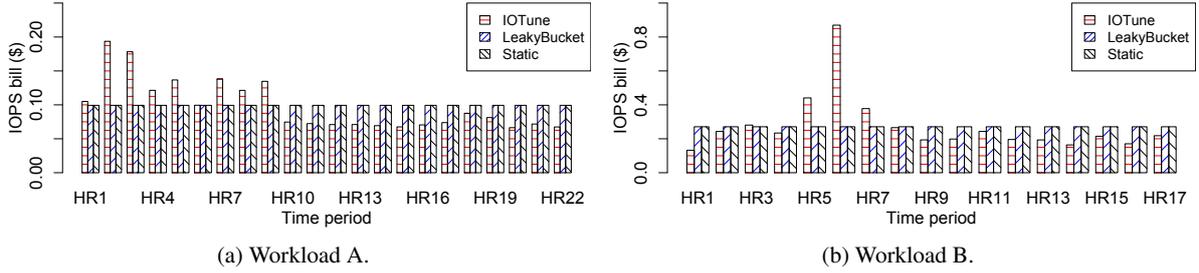

(a) Workload A.  (b) Workload B.

Figure 8: The per-hour IOPS bill with IOTune, LeakyBucket and Static resource provisioning.

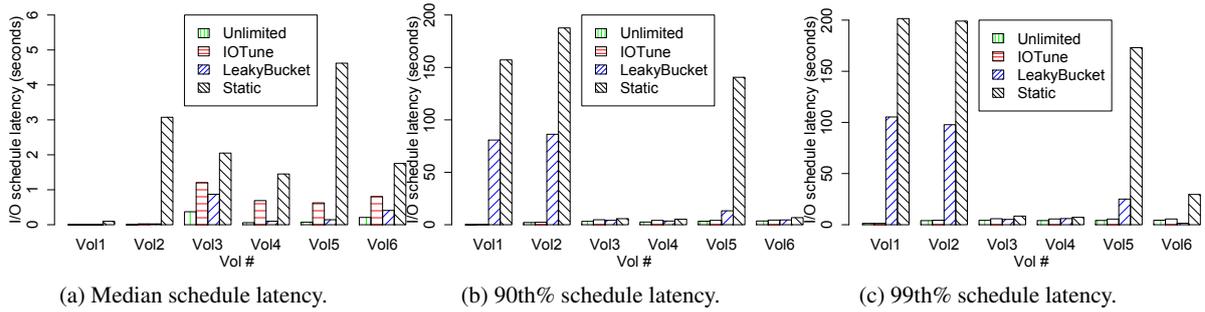

(a) Median schedule latency.  (b) 90th% schedule latency.  (c) 99th% schedule latency.

Figure 9: End-to-end I/O schedule latency with IOTune, LeakyBucket and Static resource provisioning.

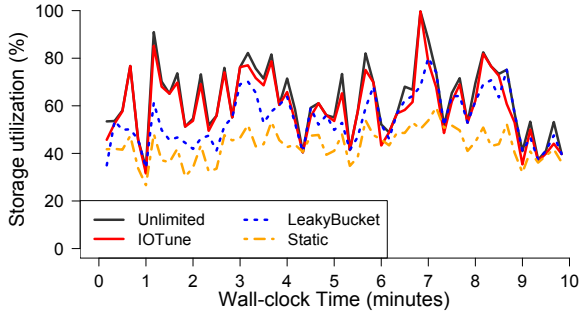

Figure 10: Storage utilization: IOTune vs. Static provisioning. We divide consumed resources by provisioned resources to calculate the utilization, which does not necessarily reflect the storage hardware utilizations.

Figure 9 demonstrates the end-to-end I/O latencies of *IOTune*, *LeakyBucket* and *Static* provisioning. For all volumes and in all cases, IOTune significantly beats the static provisioning and keeps the I/O latency in the same order of magnitude of the *Unlimited*, which imposes no IOPS limits on volumes. For volume 1, 2, 5, compared with Static provisioning, IOTune reduces the 90th% and 99th% latencies by one to two orders of magnitude. Volume 1, 2, 5 have far higher 90% and 99% latencies than volume 3, 4, 6. This can be explained by Table 2, from which we can see the former volumes have much higher 99% to 90% IOPS ratios, thus more dramatic bursts. The I/O bursting supported by LeakyBucket reduces the median latencies for workloads with moderate request rates. But for workloads with some intensive I/O periods such as Vol1 and Vol2, LeakyBucket cannot reduce the tail latency close to the Unlimited case. In contrast, IOTune can always ensure the tail latency is close to that of the Unlimited reservation policy.

For latency evaluation we assume the long I/O delays are tolerated by tenants. However, for interactive applications, users may leave if a server fail to response in seconds. For storage systems, I/O redirection is used to offload long delayed requests [12]. In these cases, storage system utilization will decrease due to I/O exodus. We assume requests are leaving if the schedule latency is higher than one second. We evaluate storage utilization of IOTune, LeakyBucket, and Static. If the initial provisioning is set at 90 percentile arrival rate, IOTune achieves 97% utilization of the Unlimited case and has 13% higher utilization than Static. The higher utilization results from the I/O intensive periods, since in low-load periods Static can also satisfy the demand and few requests will be dropped. Figure 10 presents real-time storage utilization of a ten-minutes I/O intensive period, during which IOTune delivers about 20% higher storage utilization than Static. If the initial limit is set at 80th% arrival rate, IOTune achieves 91% utilization of the Unlimited case and has 27% higher utilization than Static. LeakyBucket can improve the storage utilization considerably. But on average storage utilization achieved with IOTune is about 8% higher than the LeakyBucket policy.

## 5 Related Work

Achieving predictable performance is critical for cloud infrastructures. Isolation is a prerequisite of predictable performance. For isolation of storage performance, PARDA [31] combines a per-host flow control mechanism and a fair queuing mechanism for host-level scheduler. Per-host I/O queue size adjustments can control the I/O rate of each host to ensure host-level fairness. The VM end-to-end proportional-share fairness is achieved by using a fair queuing mechanism, which implements proportional-sharing of the host OS issue queue. mClock [32] implements VM-level proportional-share fairness subject to minimum reservations and maximum limits, which support predictable performance as well as I/O bursting. Vanguard [33] implements a full I/O path in the Linux kernel that provisions competing workloads with dedicated resources. PriorityMeister [34] employs a combination of per-workload priorities and rate limits to provide tail latency QoS for shared networked storage. vFair [35] defines a per-IO cost allocation model, which employs an approximation function to estimate the saturation throughput for combinations of various IO types of a VM. Then the saturation throughput will be combined with the share to decide the resource quota of a VM.

Another major factor for the success of the cloud is its pay per use pricing model and elasticity [36]. Elasticity enables a system to adapt to workload changes by provisioning and de-provisioning resources in an autonomic manner. Currently, infrastructure and storage elasticities are mainly achieved through adding, removing VM, container instances or storage devices [37, 38, 39, 40, 41, 42]. Morpheus [42] reprovisions the number of containers to achieve cluster-level SLO guarantee. Live data migration has been employed to achieve database elasticity [43]. For elastic storage space, Nicolae et. al. [44] propose a solution which leverages multi-disk aware file systems to hide the details of attaching and detaching virtual disks so as to circumvent the difficulty of resizing virtual disks on-the-fly. Carillon [45] enables space elasticity in storage systems via reducing and reconstructing soft state. Trushkowsky et. al. [38] propose SCADS Director, a control framework that reconfigures the storage system on-the-fly in response to workload changes to meet stringent performance requirements. Nicolae et. al. [46] propose transparently leveraging short-lived high-bandwidth virtual disks to act during peaks as a caching layer for the persistent virtual disks where the application data is stored so as to transparently boost the I/O bandwidth of long-lived virtual disks. These solutions are achieved at coarse granularities, causing heavy data movement which hurts system performance [38]. Xu et. al. propose SpringFS [39], employing read offloading and passive migration to reduce data movement in order to improve the agility of storage resizing.

It is attractive, valuable but challenging to adjust resource provisioning according to user demands in a fine-grained fashion [47]. Kaleidoscope [48] achieves elasticity by swiftly cloning VMs into many transient, short-lived, fractional workers to multiplex physical resources at a fine granularity. PRESS [49] achieves fine-grained and elastic CPU resource provisioning by adjusting the CPU limits of the target VM setting controls on the Xen credit scheduler. However, storage resource re-provisioning is challenging due to its heavy data movement. Software-defined storage policies have been implemented to enable I/O flows to achieve dynamic rate limits in network storage environments [21]. Libra [3] employs dynamic I/O usage profile to translate application-level request throughput into device-level I/O operation metrics to reserve resource in terms of application-level metrics such as key-value GET/s and PUT/s for tenants. Crystal [50] enforces a global IO bandwidth SLO on GETs/PUTs using a software-defined object storage framework. The I/O credit mechanisms enable EBS *gp2* SSD volumes [10] to burst to 3000 IOPS for extended periods of time. All these work manifest the call for in-situ elasticities of cloud storage, which is what IOTune aims to achieve.

Market-based resource allocation has been widely discussed in [51]. Pricing is another important concern in public clouds [52]. The I/O credit mechanism of EBS *gp2* SSD volumes allocates excessive resources to volumes based on credit balances. I/O credit mechanism is suboptimal for user experience as well as storage utilization. For example, when a volume demands more resources, if the volume does not have I/O credit balance, its performance cannot be bursted even if the tenant desires to pay for the burst performance. Supporting flexible user-defined QoS is a trend in cloud environments. Availability Knob [53] as well as its game theory pricing model have been proposed to support user-defined availability so as to reduce provider costs, increase provider profit, and improve user satisfaction. Recent pricing policies of cloud storage have forced tenants to migrate data among various storage options based on data access pat-terns and existing pricing policies to minimize storage costs [54]. However, data migration is usually expensive. We believe flexible pricing models directly supported by providers are desired, while in-situ elasticity requires adaptive pricing policies. Our multi-level pricing policy accompanying G-states of storage volumes provides a market-driven paradigm for block storage pricing.

## 6 Conclusion

We propose *IOTune*, a G-states driver for elastic performance of block storage. G-states enables a block device to have dynamic performance at multiple gears. IOTune utilizes software-defined storage primitives to automatically adjust the block device performance in-situ and in real time. We present the implementation of *IOTune* on an Openstack cloud platform. Tests on our staging servers verify that compared with static resource reservation the G-states feature supported by IOTune enables a volume to achieve same QoS with 50% less resource reservation, which halves volume bills. Our tests also show that compared with static IOPS provisioning IOTune reduces the end-to-end I/O tail latencies by one to two orders of magnitude. Our evaluation verifies that IOTune overcomes the disadvantages of the leaky bucket based I/O credit mechanism, which cannot handle long-period I/O bursts. In contrast, G-states supported by IOTune does not have this limitation. We also propose a new multi-level pricing policy for G-states enabled block devices. The new policy is a natural extension of current static pricing policy. Our evaluation demonstrates that G-states lower price-performance ratio of storage volumes, thus, creating values for tenants and promoting the competitiveness of providers. G-states also promote storage utilization, directly creating values for cloud providers.